# EXAMINING A HIDDEN ASSUMPTION OF BELL'S THEOREM
# AND
# COUNTEREXAMPLES TO BELL'S THEOREM
# IN THE SPACE OF ALL PATHS FOR A QUANTUM SYSTEM

Warren Leffler
Department of Mathematics,
Los Medanos College,
2700 East Leland Road Pittsburg,
CA 94565
wleffler@losmedanos.edu

ABSTRACT

This paper implements in a simple but rigorous fashion a model of particle interaction involving all paths within a quantum system, both for configuration space and for spin. The model, which we call the space of all paths, leads to a locally explicable conceptual framework for quantum mechanics. Using it we present three counterexamples to Bell's theorem. Moreover, we show that the result places severe constraints on possible viable interpretations of quantum mechanics: Either an interpretation must in some form represent a quantum system in terms of all paths within the system or, alternatively, the interpretation must harbor "action at a distance."

## I. INTRODUCTION

The philosopher Tim Maudlin has a very useful formula for characterizing various failed attempts to refute Bell's theorem: "No local *X* theory can make The Predictions [the quantum mechanical predictions] for the results of experiments carried out very far apart."[1] Assuming Bell's theorem is correct (as Maudlin argues), the X is superfluous. He calls this "The Fallacy of the Unnecessary Adjective."

The interesting thing, however, is that all proofs of Bell's theorem (his original proofs and those by others in the same vein) for two entangled particles involve a probability distribution. This means that there is a tacitly assumed "X"—namely, that the underlying space for a quantum system is measurable. In other words, if we choose "X" to be "measurable" then in Maudlin's formula we have the proposition, "No local, measurable theory can make The Predictions for the results of experiments carried out very far apart." We consider Bell's simple proof of this proposition to be obviously valid.

In this paper, however, we use path integrals to implement mathematically a model of particle interaction originally presented qualitatively by David Deutsch and described below in Sec. II. We call this the "space of all paths," SP. The space of all paths is not measurable. Indeed, it is well known that there can be no non-trivial, countably-additive measure on the space consisting of all continuous paths joining two points in coordinate space.[2] But here is a quick sketch showing that no candidate for a measure over SP can be both additive and translation invariant on the space.

Suppose there were such a measure on SP—for example, a Lebesgue measure $\mu$ on $\mathbb{R}^{\mathbb{N}} \subseteq \mathbb{R}^{\mathbb{R}}$. Let $B$ be a sphere of radius 2 in $\mathbb{R}^{\mathbb{N}}$ centered at the origin; and let $\alpha_k$ be an orthonormal basis for $\mathbb{R}^{\mathbb{N}}$. For each $k$, form the sphere $B_k$ of radius ½ centered at $\alpha_k$. Then the spheres $B_k$ form a disjoint class, and so we see that we have a contradiction (because a Lebesgue measure $\mu$ must be finite on bounded Borel sets, additive, and translation invariant):

$$\mu(B) \geq \sum \mu(B_k) = \frac{1}{2} + \frac{1}{2} + \frac{1}{2} + \ldots = \infty.$$

Countable additivity is in fact essential to all proofs of Bell's theorem for two entangled particles. Still, it is perhaps not surprising that no one has noticed this until now. Indeed, in studying quantum mechanics virtually everyone is trained initially in some form or other of von Neumann's rigorous Hilbert-space approach to the foundations. For example, here is a statement from David Griffiths's excellent introductory text: "… such non-normalizable solutions cannot represent particles, and must be rejected. Physically realizable states correspond to the square-integrable solutions to Schrödinger's equation." [3] Measurability is of course presupposed in such approaches to the foundations.

Certainly, in and of itself, lack of measurability of the underlying space does not mean that Bell's conclusion will fail to hold in the space of all paths. It only means that his particular proof, and similar ones by others for two particles, cannot be carried out in SP because of the tacitly assumed premise of measurability. In principle there could be some clever way around this. But SP does much more than present a technical difficulty to Bell's proof. The additional freedom contained in the space of all paths provides information to the two-entangled particles about the paths not taken by the original (called "tangible") particles. We shall show that it is this common information between the sides of the experimental setup (pre-existing and pre-determined following the source event) that coordinates the correlations, doing so without any causal influence being transmitted across the origin. In other words, there is no "spooky action at a distance" when widely separated measurements are performed on the respective sides. The resulting theory is thus one of local realism (in Bell's phrase, it is "locally explicable").

Of course one could reject the space of all paths as an approach to quantum mechanics, although it reproduces all the usual experimental outcomes. But in Sec. III we derive the rather remarkable result that the only alternative to some form of SP is to accept action at a distance. We establish this by extending an elementary but important argument by Bernstein, Green, Horne, and Zeilinger (BGHZ). [4]

## II. COUNTEREXAMPLES TO BELL'S THEOREM IN SP

As stated above, the SP system is a path-integral implementation of a model of particle interaction originally described qualitatively by Deutsch.[5] Deutsch based his model on Hugh Everett's relative state formulation of quantum mechanics or, as it is customarily known, MWI (the many worlds interpretation). For example, in analyzing the two-slit experiment Deutsch argued that when an ordinary particle (a *tangible* particle) travels between two points the interference effects are the result of interactions with counterpart *shadow* particles. The shadow particles interfere only with tangible particles of the same type, and therefore they can be detected only indirectly, through their effects on regular, ordinary particles—that is, shadow photons interfere only with regular photons, shadow electrons only with regular electrons, and so on. In our implementation of the SP system we fill in a one-to-one fashion each possible path in a quantum system with one of Deutsch's particles traveling that path.

Thus in our implementation of SP we postulate that whenever a single tangible particle travels from one point to another in a space (whether for position/momentum or for spin) and has a choice of paths that it can take, it *randomly* takes one of the possible paths, and distinct shadow particles take the others. We call this the "random path postulate," RPP. Thus, all possible paths that could be traveled by the tangible particle are filled in a one-to-one correspondence by accompanying shadow particles. For position/momentum the paths are in $\mathbb{R}^3$, or in a topologically constrained subspace of $\mathbb{R}^3$. For spin the paths are "rotational" paths in $\mathbb{R}^3 \times SO(3)$, where *SO*(3) is the group of rotations of $\mathbb{R}^3$ with determinant 1. We further assume that shadow particles obey the same dynamics—for example, the Lagrangian—as their tangible counterparts. For two or more non-interacting tangible particles we use the standard rule that "the amplitude that one particle will do one thing and the other one do something else is the product of the two amplitudes that the two particles would do the two things separately,"[6] which we abbreviate as NIP (for "non-interacting particles").

Hence from the perspective of SP whenever a tangible particle travels between two points, whether the "points" are in $\mathbb{R}^3$ or in $\mathbb{R}^3 \times SO(3)$, the event is associated with all possible paths between the points. We call the collection consisting of the tangible particle and its shadow counterparts the tangible particle's *shadow stream* (note that we include the tangible particle as a member of its shadow stream).

Recall that in the approach to quantum mechanics based on the Schrödinger equation one typically proceeds by finding the eigenfunctions of the quantum Hamiltonian operator (corresponding to the Hamiltonian of classical mechanics) and the associated eigenvalues. Decomposition of a state into these eigenfunctions then determines the propagator and the system's time evolution. The path-integral approach, as an alternative, is based on the Lagrange formalism of classical mechanics, with the action as the central concept. In this approach, which Feynman discovered in 1942,[7] one begins by evaluating a functional integral—the sum of exponentiated action terms over all paths in the associated quantum system—that directly yields the propagator required to determine the time-evolution dynamics of the quantum system.

The SP system is in effect a synthesis of Feynman's formulation of QM with MWI. In this system, however, one can view all particle interaction as taking place equivalently in just the world that we experience, rather than in the multi-world structure of MWI. This has several important consequences. For example, although Bell's theorem is not applicable to MWI (because MWI has multiple outcomes for each quantum event, whereas the theorem requires a single outcome) the theorem is meaningful in SP, and we shall show that it fails in the SP system. Moreover, although we do not discuss it further here, the SP system immediately eliminates many of the standard issues regarding MWI (for example, the derivation of Born's rule).[8] Finally, notice that the true statements of standard quantum mechanics plus those of SP are satisfiable in all quantum structures (such as the double-slit experiment, etc.). In other words, the two systems generate the same experimental outcomes—the same observables. Therefore by a well-known result in mathematical logic the combined theory is consistent.[9]

In standard quantum mechanics (SQM) the wavefunction is often said to contain all the information about the quantum system it represents. From the perspective of the space of all paths, however, it contains only the *extractable* information, what is left out is typically infinite, with every one of the usually infinite number of paths in SP corresponding to a distinct physical state of the system.

Indeed, we know that each distinct *preparation* procedure in SQM corresponds to an input of information to a quantum system, a distinct wavefunction—usually an uncountable infinity of possibilities. However, the output, the *extractable* information in SQM—because of its relation to orthogonal wavefunctions—corresponds to only a small fraction (indeed, only a vanishingly small fraction) of the system's *total* information in wavefunction form.

Indeed, a Hilbert space representation of a quantum system is just a homeomorphic image of SP, where homotopic classes of paths in SP are mapped onto orthogonal basis vectors in the Hilbert space. (There are examples of this below.) This is generally an infinitely-many-to-one mapping and results in an enormous loss of the quantum system's pre-existing information. As a rough analogy, it resembles the information lost in going from, say, the integers to a homomorphic system consisting of the integers mod $n$. (Actually, a better analogy would involve the real numbers mod $2\pi$, since it is this covering homomorphism in algebraic topology that maps a simply-connected space such as the real numbers down to a multiply-connected one such as spin. Incidentally, it is the fact that SQM is an infinitely-many-to-one image of SP that accounts for the intuition of some (famously including Einstein) that SQM is incomplete.

As pointed out above, Bell's elementary proof of his theorem is obviously valid when we assume that the underlying quantum system is described completely by a standard Hilbert space. Yet the proof is ruled out in SP because of the tacit measurability assumption. As mentioned, the space of all paths goes "outside the box" of the standard setup and provides, through the generated shadow streams, complementary information to the tangible particles about the paths not traveled by the tangibles.

The fundamental fact about two-particle correlations is that there is a necessary geometric symmetry between the two sides of the experimental arrangement. In the case of spinless particles, for example, a two-particle interferometer is obviously designed to have such symmetry (though we have never seen this explicitly stated). It is what underlies the so-called *principal of indistinguishability*: "If one can even in principle distinguish the path each photon has taken, then one obtains the classical probabilities (no interference)."[10] In the case of spin, however, the symmetry only becomes clear in the path integral formulation. For spin the measurements don't have to be performed equidistantly from the source (as is necessary for spinless particles), so long as there is no passage through an intervening magnetic field before the Stern-Gerlach (SG) measurements are made. One spin measurement could be carried out, say, relatively near to the source and the other measurement far away from it. For spin the symmetry comes from "rotational" paths in $\mathbb{R}^3 \times SO(3)$, as the particles on each side pass through a magnetic field, rotating along congruent spin paths: In SP tangible and shadow particles with spin are like tiny classically spinning tops, tiny dipole magnets; the path integral quantifies the net effect of particles rotating along a homotopy class of paths as the particles pass through a magnetic field.

For both spinless and spin particles, this symmetry means that there are correspondingly congruent paths on each side of a two-particle experiment. The congruent paths common to the two sides of the setup ensure that there will be correspondingly equivalent action phases for the various particles (tangible and shadow) as they travel the paths. These action phases are like clock-hands that rotate at a frequency determined by the Lagrangian as the particles travel their respective paths between points. [11, 12] The information held in the action phases is shared on each side of the experimental setup when the tangible and shadow particles come together and interfere locally with each other, determining the observed outcomes.

## A. Counterexample 1: Spinless particles in an interferometer

Our first counterexample is based on a two-particle interferometer (Rarity-Tapster[13]) for spinless particles, with its beamsplitters, mirrors, and two phase shifters $\alpha$ and $\beta$ as shown below in Fig. 1. For two-particle correlations to take place in such a setup there must be a certain amount of positional uncertainty in the source—unlike for single-particle interference, which requires a point source.[14] This means that there are infinitely many paths from the source to the beamsplitters (and thus to the detectors), handled by the mirrors.

In the next two counterexamples we focus on products of propagators over the respective homotopy classes for each side. We could do the same here, where the propagators sum over paths going to a beamsplitter, and then act on a ket representing travel to a detector. But this machinery is not necessary in this elementary structure, and to simplify the discussion we can view each homotopy class as consisting of just a single path, so that there are just four particles in all for our setup, two tangible particles and two shadow particles. (Actually, what we are doing here is working in a particular homeomorphic image of the richer SP space, mentioned above.) This enables us to carry out our analysis in the conventional fashion using Dirac notation familiar to most readers.

Thus in our setup there are four homotopy classes of paths: $a$, $a'$, $b$, and $b'$.

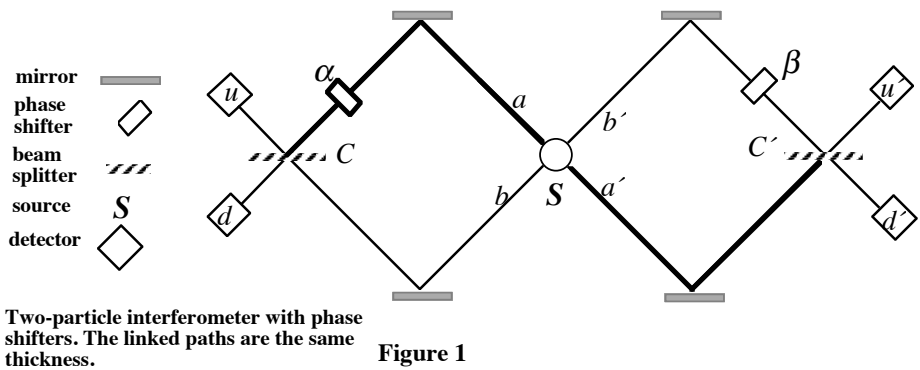

Two-particle interferometer with phase shifters. The linked paths are the same thickness.

Figure 1

A more important and necessary condition, however, is that the emission angle subtended at the source by the paths of the two entangled tangibles must be sufficiently acute for correlations to occur (if not—if the angle is too large—then, as is well known empirically, each side corresponds to single-particle interference, and the correlations disappear).[15] This means that there are just two choices for what we term the "upper" tangible particle (recall that for simplification we consider each homotopy class as consisting of a single path): The upper tangible can only take an upper path—that is, either $a$ or $b'$ in the figure—which it does randomly at each trial of the experiment. If the upper tangible takes, say, path $a$ then the other, lower tangible must by conservation of momentum take the linked path $a'$. Similarly if the upper takes path $b'$ then the other tangible is linked to take path $b$. The linkage is indicated by the same thickness of the lines in the figure. Moreover, note that the particles taking paths $a$ and $b$ are from different streams; likewise for those taking paths $b'$ and $a'$.

We can now express the probability that the tangible particles are both observed at the upper detectors $u$ and $u'$, when measurements (phase shifts) of $\alpha$ on the left and $\beta$ on the right are made (a similar argument works for $d$ and $d'$). We do this by normalizing and taking the absolute square of the following amplitude given in Dirac notation.

$$\langle u|a\rangle\langle u'|a'\rangle + \langle u|b\rangle\langle u'|b'\rangle \qquad (2)$$

Now the amplitude in expression (2) is just the usual SQM expression for particles in the singlet state. In Schrödinger's definition, it comes from an "entangled state," one that cannot be factored into states coming separately from the tensor product spaces. Thus, so far there is nothing new. Also, as it stands, expression (2) seems to require that the particles must somehow communicate across the source in order to coordinate the predicted correlations. Of course this is exactly what Bell's theorem purports to prove—that the correlations are "*locally inexplicable*."[16]

But wait! Here is where shadow particles make all the difference: Note that $\langle u'|a'\rangle = \langle u|b\rangle$. This is clear from the geometric symmetry of Fig. 1, where we are assuming that on each side shadow particles are simultaneously traveling in one of the two homotopy classes while tangible particles are traveling in the other. Thus, "substituting equals for equals," we see that expression (2) is equivalent to

$$\langle u|a\rangle\langle u|b\rangle + \langle u'|a'\rangle\langle u'|b'\rangle. \qquad (3)$$

In other words, we have "factored" the information in expression (2) by means of information carried by shadow particles on each side of the experiment, where either the tangible travels along an upper path and its shadow counterpart travels the lower one, or vice versa (keeping in mind our simplifying assumption about homotopy classes). Indeed, the information in each addend of expression (3) is now local to the corresponding side of $S$ ($a$ and $b$ on the left, $a'$ and $b'$ on the right). Moreover, since the particles traveling along $a$ and $b$ (also along $a'$ and $b'$) come from different streams, this mirrors the actual physical situation. In other words, the composite amplitude according to NIP (Sec. II) is exactly this sum of the products of the separate amplitudes, $\langle u|a\rangle\langle u|b\rangle$ and $\langle u'|a'\rangle\langle u'|b'\rangle$. We now sketch the standard and well-known calculation that leads to the usual probability of the correlations.

But first we note that our simplifying assumption above of identifying a small finite number of paths (four) with the homotopy classes containing them (the homotopy classes having infinite cardinality) allows us to avoid the path integral in our calculation, although it is in effect actually behind our assigning amplitudes to the paths in SP. But when we treat spin correlations in one of our following counterarguments below, path-integral complexity and the propagator (time development operator) inextricably enter into the discussion. For present purposes, however, a simple conventional procedure using Dirac notation for amplitudes will do nicely.

Indeed, recall that the amplitude of an event in QM is just a complex number of the form $re^{i\theta}$, where $r$ is real.[17] Since the lengths of the upper paths are the same (except for the phase shifters), as are those of the lower ones, we let $r_1 e^{i\theta}$ and $r_2 e^{i\theta'}$ be the exponentiated action terms denoting the upper and lower path amplitudes. Also we use the standard factor of $i$ ($= e^{i\pi/2}$) for reflection at a beamsplitter. Thus, incorporating the measurements indicated by phase shifts on each side, we have,

$$r_1 e^{i\theta} e^{i\alpha} i = \langle u|a\rangle, r_2 e^{i\theta'} = \langle u'|a'\rangle = \langle u|b\rangle, r_1 e^{i\theta} e^{i\beta} i = \langle u'|b'\rangle. \qquad (4)$$

This gives,

$$\langle u|a\rangle\langle u|b\rangle + \langle u'|a'\rangle\langle u'|b'\rangle =$$
$$(r_1 e^{i\theta} e^{i\alpha} i)(r_2 e^{i\theta'}) + (r_2 e^{i\theta'})(r_1 e^{i\theta} e^{i\beta} i) = i \cdot r_1 r_2 e^{i(\theta+\theta')}(e^{i\alpha} + e^{i\beta}) \qquad (5)$$

Hence, because of the similar amplitudes contained in expression (2), when we take the absolute square of Eq. (5) (that is multiply the right side of Eq. (5) by its complex conjugate) we obtain a result proportional to

$$(e^{i\alpha} + e^{i\beta})(e^{-i\alpha} + e^{-i\beta}). \qquad (6)$$

A similar calculation yields the same result for both particles registering at the lower detectors. Combining the result for the lower detectors with that of expression (6) for the upper detectors, and inserting appropriate normalizing constants, we have (by the elementary arithmetic for complex numbers) the probability that both upper detectors or both lower detectors will fire:

$$\cos^2 \frac{\alpha - \beta}{2}. \qquad (7)$$

This is the well-known probability of the correlations. In SQM an amplitude expression is completely opaque. It masterfully yields the quantum predictions but (unlike SP) reveals nothing about the underlying mechanism—which, as we saw, hinges on shadow particles traveling paths not taken by the tangibles. The correspondingly congruent paths between the two sides lead to the observed correlations. Where the tangibles are ultimately seen when they arrive in the regions of the detectors is an outcome of interference effects determined by a contiguous exchange of information between the tangible particles and their shadow counterparts on the respective sides. Nothing needs to be sent across the origin. There is no collapse of a wavefunction. The sharing of information between the tangible and shadow particles occurs at the meeting points on each side ($C$ and $C'$ in Fig. I). The result is an outcome of an entirely locally explicable process—via local interference effects. There is no action at a distance, although the end result is as though there were. **QED**

In a moment we shall carry out two similar arguments, one for entangled quantum beads on a ring, and one for spin. In a strict logical sense these are unnecessary, since it only takes one counterexample to overturn a theorem. Still, for historical reasons, the argument for spin has some interest in its own right, because in the original 1964 proof of his theorem Bell had in mind David Bohm's spin version of a gedanken experimental test of the EPR correlations.[18] The argument for entangled beads (the next counterexample) prepares the way for the spin argument.

A further comment: As we noted above, given the simplicity of the interferometer setup we never bothered to refer to amplitudes such as $\langle u|a\rangle$ and $\langle u|b\rangle$ in expressions (2) and (3) as "propagators" over homotopy classes. But that is what they involve: They determine the time evolution of particles going to a detector when the particles travel in homotopy classes of paths meeting at a beamsplitter. (More properly, the propagator is the sum of amplitudes for particles traveling to a beamsplitter. It acts on the ket corresponding to passage from the beamsplitter to a detector.)

In the somewhat more complicated context of the following counterexample, and also the next one, we will find it helpful to carry out our argument explicitly in terms of "propagators" (ignoring the ket corresponding to a measurement at a detector).

## B. Counterexample 2: Entangled ring particles

In elementary quantum mechanics a frequently treated example is the so-called quantum bead-on-a-ring (roughly pictured by an electron moving on a conducting ring in, say, benzene). Lawrence Schulman (as a preliminary to presenting his path integral for spin)[19] gave (so far as we know) the first-ever path-integral solution to this problem, a problem that is easily solved in the traditional approach using Schrödinger's equation (taking momentum to be the Fourier transform of position, etc.). Using either approach, one shows that there is a discrete set of observables—for example, a denumerably infinite set of eigenvalues for momentum (in contrast to what occurs with a classical bead traveling around a frictionless ring where, in principle, we can obtain any of a continuum of momentum outputs).

The associated configuration space for the ring—unlike that for Euclidean space, where Feynman's formulation was originally developed—is not simply connected, a consequence of the topological constraint of confining the tangible particle's path to a circle. This means that the overall path-propagator is a linear combination of propagators over each homotopy class in the space.[20] Thus assume that for the ring we have a free tangible particle constrained to move on a circle, the particle able to go any number of times back and forth along the circle. The homotopy classes of paths in this context correspond to winding numbers, $n$—that is, the number of times the particle travels past a given fixed point in the positive direction (counterclockwise) minus the number of times it passes the point in the minus direction (clockwise).

The propagator for the ring over a fixed time interval and for $0 \leq \alpha, \theta \leq 2\pi$ is

$$K(\alpha,\theta) = \sum_{n=-\infty}^{\infty} K_n(\alpha,\theta) \qquad (8)$$

where $K_n(\alpha,\theta)$ is proportional to the sum of all exponentiated action terms for paths around the circle in the homotopy class corresponding to the winding number, $n$ (the action for the propagator is just that of a free particle).[21]

Now—in an instructive but doubtless experimentally infeasible gedanken experiment—consider a pair of entangled particles emitted by a source event and traveling in opposite directions along a pair of rings that are somehow separated and moved apart following the source event. We will assume that the particles receive the same initial momentum kick, starting from some initial point $\theta$ on the associated initial pair of circles. The two tangible particles will then, randomly, travel in a pair of corresponding homotopy classes (by conservation of momentum), which we denote by $n$ and $-n$. This is a local event occurring at the source, a consequence of the initial input of momentum. Assume now that a momentum measurement at angle-point $\alpha$ is made on the left circle and one at $\beta$ is made on the right. By RPP and NIP above the combined amplitude for this is

$$\sum_{n=-\infty}^{\infty} K_n(\alpha,\theta) K_{-n}(\beta,\theta) \qquad (9)$$

The propagators in the sum in expression (9) contain the information that gives the time-evolution of the quantum system for the entangled ring particles.

In one of several cases involving the relative positions of the angle-points on the two circles, assume that $0 \leq \theta \leq \beta \leq \alpha < \pi$ (the arguments for the other cases are similar). Now it is already intuitively clear that the propagator $K_n(\alpha,\theta)$ dealing with the left side of the experimental arrangement contains the information of the propagator $K_{-n}(\beta,\theta)$ on the right side, since the set of all paths from $\theta$ to $\alpha$ includes those from $\theta$ to $\beta$. Thus the product of amplitudes in expression (9) is "almost completely" based on paths that on each side of the experimental setup are congruent to corresponding ones on the other side. But a simple mathematical argument below brings out the correspondence more precisely.

Consider any path from $\beta$ to $\alpha$ of winding number 0. This path, in traveling from $\beta$ to $\alpha$, can loop any number of times in a given direction around the reference point 0 on the circle, but it must always in its overall passage around the circle travel (but not in any particular sequence) the same number of times past 0 in the opposite direction. Hence, for each $n$, any path from $\theta$ to $\alpha$ of winding number $n$ is homotopically equivalent to one from $\theta$ to $\beta$ of winding number $n$ plus one from $\beta$ to $\alpha$ of winding number 0.

As is well known, the exponentiated action for a free particle traveling around the circle on any path from $\beta$ to $\alpha$ of winding number 0 is a function $f(\alpha,\beta)$ whose value is proportional to $\exp\left(\frac{im}{2\hbar t}(\alpha-\beta)^2\right)$.[22] One can see this intuitively, since the exponentiated action for paths corresponding to congruent but oppositely directed paths around the circle from $\beta$ to $\alpha$ will cancel, leaving just the action over the minimal path from $\beta$ to $\alpha$. Thus we have

$$\sum_{n=-\infty}^{\infty} K_n(\alpha,\theta)K_{-n}(\beta,\theta) = f(\alpha,\beta) \sum_{n=-\infty}^{\infty} K_n(\beta,\theta)K_{-n}(\beta,\theta). \tag{10}$$

The amplitude for finding correlated momentum observables thus involves a function that depends locally on the least path from $\beta$ to $\alpha$. **QED**

### C. Counterexample 3: Entangled spin1/2 particles

We shall reduce the argument for this counterexample to that of the previous one for ring particles in the following way:

First, note that the paths for spin in SP are "rotational" paths in $\mathbb{R}^3 \times SO(3)$. As is well known, we can model $SO(3)$ homeomorphically in terms of a solid ball of radius $\pi$, with antipodal points identified. In this picture, a point a distance $\gamma$ from the center of the ball and along the radius pointing in the direction of the unit vector $\mathbf{n}$ corresponds to a counterclockwise rotation $\gamma$ about the axis $\mathbf{n}$. Because $SO(3)$ has only two homotopy classes of paths (its fundamental group is of order 2) and because we are identifying antipodal points on the ball, it is an interesting fact that a path along, say, the y-axis that passes just once through the outer surface of the ball is not homotopic to a path that does not pass through at all, although paths that pass through any even number of times are so homotopic. To see this, suppose a path along the y-axis runs from (0, 0, 0) (the center of the ball) in the positive direction to the point (0, $\pi$, 0) on the ball) and then jumps back to the antipodal point (0, $-\pi$, 0), and continues on to do this a second time, so that it runs from 0 to $4\pi$ along the y-axis. This generates a closed loop in which the paths can be moved continuously to the ball's surface, still connecting $(0, \pi, 0)$ to $(0, -\pi, 0)$ twice. Now mirror the second half of the path to the antipodal side, which produces a closed loop on the surface of the ball connecting the point $(0, \pi, 0)$ on the *y*-axis to itself along a great circle. This circle can be shrunk to a point. We will take this into account below, when we define a kind of winding number for paths looping back and forth along the *y*-axis.

Now consider a pair of entangled spin-1/2 (tangible) particles traveling in opposite directions along, say, the *y*-axis to SG devices on each side of an experimental setup. As we have emphasized, in SP a path is a continuous sequence of rotations in the space *SO*(3) (that is, in $\mathbb{R}^3 \times SO(3)$). Now assume, for convenience, that both tangibles after they leave the source event are passed through a filter oriented at angle $\theta$ along the *y*-axis. Assume also that a measurement of angle $\alpha$ is made on the left side about the *y*-axis, while one of $\beta$ is made on the right, where (as in counterexample 2) $0 \leq \theta \leq \beta \leq \alpha < \pi$. Thus, as noted in the discussion of expression (8) above, the propagator corresponding to paths from $\theta$ to $\alpha$ on the left side of the experimental setup contains the information of all the paths going from $\theta$ to $\beta$ on the right, since the set of all paths from $\theta$ to $\alpha$ includes those from $\theta$ to $\beta$. Moreover, we shall conclude as follows (as in counterexample 2) that there is a formula expressing this information.

In computing the propagator from $\theta$ to, say, $\gamma$ along the *y*-axis in the ball of radius $\pi$ (that is, a rotation of angle $\gamma$ about the *y*-axis) the path integral sums the exponentiated action terms over *all* possible paths from $\theta$ to $\gamma$ in the ball. But, in a key simplification for our discussion, we can assume dominance of rotations about the *y*-axis (so what we have here is an approximation). Thus assume that the SG field is titled at some angle about the *y*-axis and that the tangible on the left side is traveling in the *SO*(3) homotopy class *A* and the other tangible in *A´* on the right.

Hence, on each side we picture the tangible's path as being projected onto a continuous path along the *y*-axis of the ball. In going from $\theta$ to another angle-point $\gamma$, such a path *confined* to the *y*-axis, can travel any number of times back and forth, sometimes passing through the ball. This will enable us to carry out a construction similar to that of the ring counterexample above, but where we can ignore the difference in the actions involved in the two contexts. (As mentioned, the action for spin is the sum of a spinning top component and a magnetic field component, but this detail will not affect our argument).[23] Also, to avoid the fact noted above about paths passing through the ball an even number of times etc., we constrain the paths to the *y*-axis. Thus a path is "stuck" to the *y*-axis and can no longer be moved continuously from the *y*-axis up to the surface of the ball, which eliminates any homotopy involving an even number of passages through the surface. This enables us to speak of winding numbers *n* about the reference point 0 for a path that goes from $\theta$ to $\gamma$ and possibly passes through the ball as the path runs back and forth along the *y*-axis.

By this construction of projecting onto the *y*-axis in the ball (corresponding to our SG measurements about the *y*-axis on each side of the experimental arrangement) we thus generate a set of homotopy classes different from the conventional two given by *SO*(3) (the usual two can be pictured by two regions on a two-sphere [24]). This enables us to reduce everything to the previous counterexample. Therefore when we sum over these homotopy classes labeled *n* we obtain a result similar to EQ (10), where the difference between the sides is just a function of $\alpha$ and $\beta$. For spin this means that the common information between the sides comes from correspondingly congruent stream paths. Thus, as with the previous two counterexamples, congruent paths on each side of the experiment lead to the correlations when SG measurements are performed on the spatially separated sides. The correlations occur by means of initial information imparted at the source to the shadow streams—that is, the correlations are predetermined at the source as a function of the subsequent measurements. The interactions producing the correlations are purely local throughout the experiment. Nothing needs to be sent across the origin. **QED**

There are of course other versions of Bell's theorem. For example, there is a three-particle version—which, at least on the surface, does not appear to involve measurability. [25] The three-particle version does, however, hinge on the notion of "an element of reality." Although an analysis of the three-particle case is beyond the scope of the present paper, we point out that SP's element of reality (because it includes the stream of shadow particles accompanying the tangible) is a rather different concept than the conventional one, and it must now be taken into account in carrying out any Bell-like argument. In SP a tangible particle is just a single particle in a stream of counterpart particles (usually infinitely many), the stream moving along according to the dictates of Schrödinger's equation, infinitely many configurations of the stream leading in general to the same observable. Moreover, as in the two-particle examples, the overall experimental arrangement for three particles must again possess a geometric symmetry—leading to congruent paths and common information shared by all three streams. In any case, by virtue of our two-particle counterexamples, all forms of Bell's theorem are now suspect.

### III. CONCLUSION

Now one can always reject shadow particles by arguing that they can be known only indirectly. But interestingly it turns out that either we must choose shadow particles or, alternatively, "spooky action at a distance." These are the two mutually exclusive choices for quantum reality (in a way perhaps the most important result in this paper). This conclusion is based (somewhat ironically given the original intent of the authors) on an obvious and straightforward extension of an argument by Bernstein, Green, Horne, and Zeilinger (BGHZ). [26]

In their paper BGHZ proved what they called a "Bell theorem without inequalities for two spinless particles." For their proof to go through, they found it necessary to augment their assumptions with a premise called "emptiness of paths not taken," EPNT. As they stated,

> … one can deny EPNT and thereby imagine that *something* could travel down the empty beam, so as to provide information to the nonempty beam, when the two beams meet. And this something could be consistent with EPR locality, if the particles (and these somethings) on opposite sides of the origin do not communicate.

Of course in SP we deny EPNT by filling the "empty" paths in a one-to-one correspondence, with each possible path being traveled by a particle, tangible or shadow. The otherwise empty paths (now filled) do indeed provide complementary information when the paths meet prior to or during measurement, and it is precisely this shared information that makes locally explicable the observed correlations for widely separated measurements in two-particle experiments. There is no need for information to be communicated across the origin in order to coordinate the correlations. But BGHZ proved that if the possible paths not traveled by the entangled tangible particles are actually empty, then the quantum correlations must occur through "action at a distance."

BGHZ based their argument on a two-particle interferometer, but their proof can be extended to any multiply-connected quantum system (the two-particle interferometer in Sec. IIA is of course an elementary example of such a system; and spin spaces in SP are also multiply connected). Indeed, by assuming that the paths not traveled by the tangible particles are empty then the straightforward BGHZ argument applies, *mutatis mutandis*, to homotopy classes of paths in a general space in place of the small number of paths analyzed by the BGHZ paper. In other words, we have the result that "SP" implies "no action at a distance," and "not SP" implies "action at a distance."

Now one could deny EPNT and still have a nonlocal theory (for example, David Bohm's pilot-wave theory, which in effect "fills" the paths with a "pilot wave").[27] Also there are partial forms of SP that claim to be local or realistic theories—for example, "consistent histories"(CH).[28] CH is a mathematically sound theory based on SQM (modifying an approach to quantum logic developed by Birkhoff and von Neumann).[29] But in CH such terms and concepts as "and," "or," and "local" no longer have their ordinary, commonsense meanings unless they are used according to CH's "single framework rule." This rule allows one to employ standard Boolean logic and concepts such as "local" only on single frameworks or histories (path sequences that do not involve non-commuting quantum operators along the sequence). This rules out as meaningless (in the sense of "speak no evil") any discussion of what appear, from a commonsense standpoint, to be the nonlocal experimental outcomes implied by Bell's theorem. Yet the single framework rule is in fact merely a grammatical or syntactical rule, although the grammar is based on the properties of operators in the Hilbert space of SQM. Indeed this syntactical rule leads to a picture of reality in which infinitely many separate slices or histories can be true, although for infinitely many of the pairs of histories taken together, any comparison between them would be meaningless.[30] CH maintains that this paradox is not a defect in the theory, but merely reflects the way the quantum world happens to be.

Perhaps, but there are no such paradoxes associated with SP. In SP "locality" has its usual meaning of contiguous interaction, and everything is an elegant sum of classical dynamics over the paths in a shadow stream. We can freely combine logical statements within the theory using classical Boolean logic. Thus like special relativity, for example, the space of all paths is straightforward and easy to grasp and conforms perfectly with informed commonsense (that is everything follows logically from the fundamental properties). Indeed, as stated earlier, the standard Hilbert-space representation of a quantum system—with its corresponding need for wavefunction collapse and decoherence—is just a many-to-one homeomorphic image of the richer SP representation when homotopy classes are mapped to orthogonal basis vectors in the Hilbert space.

Note also, contrary to what is commonly supposed in these matters: One cannot reject the space of all paths on the basis of Occam's razor ("entities should not be multiplied beyond necessity"). The reason is that Occam's razor (as stated) is vague and imprecise about the criteria for its strictures. It can, however, be expressed precisely in complexity theory and given a form amenable to mathematical analysis: "If presented with a choice between indifferent alternatives, then one ought to select the simplest one."[31] Put this way, the alternatives here are decidedly not indifferent: The non-existence of shadow particles leads to paradox and, in the case of entangled particles, to something close to an absurdity that not only conflicts with special relativity but would also have to take place according to utterly incomprehensible dynamical laws.

Now shadow particles and their bizarre implication of infinitely many parallel patterns of matter (that is, parallel copies) interacting with our tangible universe may be novel and at first seem contrary to our everyday experience, but they are not absurd: Outcomes in SP are simply a sum over classical paths, where all particles in a stream obey classical dynamics. Of course shadow particles can be known only indirectly at present; but, as we know, this has happened before in science. For example, as far back as 60 BC the Roman poet Lucretius (in his amazingly prescient poem *On the Nature of Things*) hypothesized the existence of atoms on the basis of what we now call Brownian motion—in the form of the bombardment of "dancing" dust particles by "hidden" atoms. And later, in the nineteenth century long before more direct experimental confirmation was feasible, many physicists (again thinking in terms of the indirect evidence generated by Brownian motion) conjectured the existence of atoms and molecules. Something similar is occurring with Deutsch's shadow particles. To reject them is to embrace action at a distance—which in Einstein's words "no reasonable definition of reality could be expected to permit …").[32]